\documentclass[12pt]{article}
\usepackage{epsfig}
\begin{document}

\title{Signals of $Z'$ boson in the Bhabha process within the LEP 2  data set}
\author{A.V. Gulov, V.V. Skalozub}
\maketitle

\begin{abstract}
The LEP2 data set on the Bhabha process is analyzed with the aim
to detect the signals of the heavy virtual $Z'$ gauge bosons. The
state interacting with the left-handed standard-model doublets and
called the Chiral $Z'$ is investigated. This particle was
introduced already as the low-energy state allowed by the
renormalizability of the model. The contribution of the Chiral
$Z'$ state to the Bhabha process is described by two parameters:
the coupling to electrons and the $Z$-$Z'$ mixing angle. The
sign-definite one-parameter observable is proposed to measure the
$Z'$ coupling to the electron current. The one-parameter fit of
the data shows no signals of the particle. The alternative
two-parameter fit of the differential cross-sections is also
performed. It also shows no Chiral $Z'$ signals. The comparisons
with other fits are discussed.
\end{abstract}

\section{Introduction}

Searching for new heavy particles beyond the energy scale of the
standard model (SM) is an important problem of modern high energy
physics. In particular, in recently finished experiments at LEP
either model dependent or model independent searches for different
heavy virtual states were carried out and the masses and couplings
of numerous particles entering a number of popular models
extending the SM have been restricted \cite{EWWG}. As a general
observation it was found that the low bounds on the mass of a
specific particle are strongly model dependent ones and vary in a
wide energy range. For instance, the lower limit on the mass of
the $Z'$ gauge boson changes from $670$ GeV to $2$ TeV
\cite{EWWG}. At the accelerators of the next generation (the LHC,
in particular) the signals of new heavy particles are expected to
be clearly detected and some of them could be discovered. But the
underlying theory beyond the SM will hardly be settled. These
circumstances may serve as a motivation for the model-independent
description of new heavy virtual particles with specific quantum
numbers.

In Refs. \cite{EPJC00,PRD00} the approach to pick out uniquely the
$Z'$ gauge boson in different scattering processes at low energies
has been developed and applied \cite{YAF04,PRD04} to analyze the
data of the LEP2 experiments. It is based on two main principles:
1) the renormalizability of an underlying model extending the SM
is assumed; 2) the kinematics features of the specific processes
in which the $Z'$ appears as a virtual state are accounted for.
The former requirement results in the specific relations between
the low energy couplings of the $Z'$ with the particles of the SM
called the renormalization group (RG) relations and decreasing a
number of independent parameters. The latter gives a possibility
to introduce the observables which uniquely determine this state
in different scattering processes. It is important to notice that
at each stage of this analysis one also has to take into
consideration the derived correlations. The detailed analysis of
the existing in the literature \cite{diffcs} experimental data for
the annihilation $e^+ e^- \to \mu^+ \mu^-, e^+ e^- \to \tau^+
\tau^- $ processes and the Bhabha process, $e^+ e^- \to e^+ e^- $,
having as a goal searching for signal of the Abelian $Z'$ boson,
showed \cite{YAF04,PRD04}: 1) In the annihilation processes, the
present day data are in accordance with the $Z'$ existence, but
the data set is not bulk enough to detect the signals at more than
the $1\sigma$ confidence level (CL). 2) In the Bhabha process, the
signals can be determined at the $2 \sigma$ CL. The mass of the
particle was estimated to be $m_{Z'}\sim 1-1.2$ TeV. So one has to
expect that the increase of the data set could make the signals
more evident.

An important finding in Ref \cite{EPJC00} was the result that at
low energies there are two types of the $Z'$ interactions with the
SM fermions compatible with the RG equations. The first is the
well known Abelian $Z'$ boson. And the second one is the ``Chiral
$Z'$'' interacting with the left-handed species, only. The RG
relations are different for these cases. Correspondingly,
different observables must be constructed in order to pick out the
signals of these states. So, searching for the Chiral $Z'$ is of
interest as a possibility of new physics beyond the SM.

In the present paper the effects of the Chiral $Z'$ boson in the
Bhabha process $e^- e^+\to e^- e^+$ are investigated. We introduce
the one-parametric observable to pick out the low energy coupling
of the $Z'$ to the left-handed electrons. Because of a few
independent couplings entering the cross-section there is other
interesting possibility of two-parameter fits. We find that in
both cases the LEP2 data show no signals of the Chiral $Z'$ boson.

\section{$Z'$ couplings at low energies}

To be self-contained, let us remind the necessary results on the
description of the $Z'$ interactions at low energies. Suppose that
all new heavy particles are decoupled and the $Z'$ boson is
associated with some low-energy gauge subgroup. Then, as it was
shown in Ref \cite{EPJC00}, its couplings to the SM particles
satisfy so-called RG relations required by the renormalizability
of the underlying theory beyond the SM. By accounting for the
relations the number of independent parameters describing
cross-sections can be significantly reduced. Notice that all
popular $SO(10)$-based $Z'$ models are embedded into the mentioned
relations.

To describe the types of the $Z'$ boson predicted by the RG
analysis let us introduce the phenomenological parametrization of
the low-energy $Z'$ couplings to the SM particles \cite{EPJC00}.
The couplings to the SM fermion left-handed doublets and
right-handed singlets $f$ are described by the Lagrangian
\[
{\cal L}_f=
\frac{\tilde{g}}{2}{Z}'_{0\mu}\sum\limits_{f}\bar{f}{\gamma^\mu}
     \tilde{Y}(f)f,
\]
where ${Z}'_{0\mu}$ denotes the eigenstate of the $Z'$ gauge
group, $\tilde{g}$ stands for the charge corresponding to the
group, and $\tilde{Y}(f)$ are the unknown generators
characterizing the model beyond the SM. The generators are
diagonal $2\times 2$ matrices for the fermion doublets and numbers
for the fermion singlets. In the same manner, the $Z'$ couplings
to the scalar doublet $\phi$ are derived by adding the appropriate
term to the standard electroweak covariant derivative:
\[
{\cal L}_s=\left|\left( D^{{\rm ew,} \phi}_\mu -
  \frac{i\tilde{g}}{2}\tilde{Y}(\phi){Z}'_{0\mu}
  \right)\phi\right|^2,
\]
where the generator $\tilde{Y}(\phi)$ is the diagonal $2\times 2$
matrix.

The RG relations predict two types of the $Z'$ boson -- the
Abelian and the Chiral one. The couplings of the Abelian $Z'$
boson is restricted by the relations
\begin{equation}\label{1}
\tilde{Y}(f_L)=\tilde{Y}_{fL}\cdot\hat{1},\quad
\tilde{Y}(\phi)=\tilde{Y}_{\phi}\cdot\hat{1},\quad
\tilde{Y}(f_R)-\tilde{Y}_{fL}=2T_{3,f}\tilde{Y}_\phi,
\end{equation}
where $\hat{1}$ is the $2\times 2$ unit matrix, and $T^3_f$ is the
third component of the fermion weak isospin. As is seen, the set
of independent couplings consists of one number for each SM
doublet ($\tilde{Y}_{fL}$ and $\tilde{Y}_\phi$). Introducing the
$Z'$ couplings to the vector and axial-vector currents,
\begin{eqnarray}\label{2}
&&v_f=\frac{\tilde{Y}(f_R)+\tilde{Y}_{fL}}{2}=\tilde{Y}_{fL}+T_{3,f}\tilde{Y}_\phi,
\nonumber\\ &&
a_f=\frac{\tilde{Y}(f_R)-\tilde{Y}_{fL}}{2}=T_{3,f}\tilde{Y}_\phi,
\end{eqnarray}
we see that the $Z'$ couplings to the axial-vector currents have
the universal absolute value, related to the $Z'$ coupling to
scalars. So, we can also use as an independent set of parameters
the $Z'$ couplings to the vector fermion currents ($v_f$) and the
$Z'$ coupling to the electron axial-vector current ($a_e=a$).

The second possibility is the Chiral $Z'$. It is characterized by
the constraints
\begin{equation}\label{3}
\tilde{Y}(f_L)=-\tilde{Y}_{fL}\cdot\hat\sigma_3,\quad
\tilde{Y}(f_R)=0,\quad
\tilde{Y}(\phi)=-\tilde{Y}_{\phi}\cdot\hat\sigma_3
\end{equation}
where $\hat\sigma_3$ is the third Pauli matrix. The Chiral $Z'$
interacts with the SM doublets only that can be described by one
parameter for each doublet ($\tilde{Y}_{fL}$ and
$\tilde{Y}_\phi$).

The $Z'$ couplings to the SM scalars necessarily leads to the
mixing of physical $Z$ and $Z'$ states. It is given by the
relation \cite{EPJC00}
\begin{equation}\label{4}
\theta_0 = \frac{\sin\theta_W\cos\theta_W}{\sqrt{4\pi\alpha}}
\frac{m^2_Z}{m^2_{Z'}} \tilde{g}\tilde{Y}_\phi
+O\left(\frac{m^4_Z}{m^4_{Z'}}\right)
\end{equation}
for both  the Abelian and the Chiral $Z'$ bosons, where $\theta_W$
is the SM value of the Weinberg angle. As estimates fulfilled at
LEP2 energies showed, in spite of the smallness, the $Z$-$Z'$
mixing affects the $Z$-boson couplings and can produce the effects
of the order of that induced by the $Z'$ exchange diagrams. So it
cannot be neglected.

Due to a reduced number of independent couplings it is possible to
pick out the $Z'$ signals in leptonic processes and efficiently
constraint the parameters of this particle. In our previous papers
\cite{PRD00,YAF04,PRD04} we have analyzed the signals of the
Abelian $Z'$ for $e^-e^+\to e^-e^+$, $e^-e^+\to \mu^-\mu^+$, and
$e^-e^+\to \tau^-\tau^+$ processes. We constructed the
one-parametric sign-definite observables responsible for $a^2$
coupling for each of the processes. It was also possible to
provide the one-parametric sign-definite observable for $v_e^2$ in
the Bhabha process $e^-e^+\to e^-e^+$. The LEP2 data on $e^-e^+\to
\mu^-\mu^+, \tau^-\tau^+$ processes show the signal for $a^2$ at
the 1$\sigma$ CL. The Bhabha process is mainly sensitive to the
vector coupling $v_e^2$, for which the signal is found at the
2$\sigma$ CL.

Below we turn to the analysis of the Bhabha process with the aim
to search for the Chiral $Z'$ gauge boson.

\section{The differential cross-section}

In four-fermion scattering processes $e^-e^+\to f\bar{f}$ there
are two leading-order (in the improved Born approximation) effects
related to the $Z'$-boson existence. The first one is the
amplitude with the virtual $Z'$ exchange $e^-e^+\to Z'\to
f\bar{f}$. In the lowest order in $m_{Z'}^{-2}$ it is resulted in
four-fermion contact couplings, which are various quadratic
combinations of $\tilde{g}\tilde{Y}_{f}/m_{Z'}$ for all possible
generators $\tilde{Y}_{f}$ for the considered type of the
$Z'$-boson.

The second effect is originated by the $Z$-$Z'$ mixing. The mixing
affects the $Z$-boson exchange diagram, $e^-e^+\to Z\to f\bar{f}$,
by terms of order $\theta_0\tilde{g}\tilde{Y}_{f}$. Since the
mixing angle is proportional to $m_{Z'}^{-2}$, in the amplitude
the leading-order terms are described by the products
$\tilde{g}^2\tilde{Y}_{f}\tilde{Y}_\phi m_{Z'}^{-2}$ for all
generators $\tilde{Y}_{f}$.

In what follows we will use the dimensionless coupling constants
\[
\bar{l}_f=\frac{m_Z}{\sqrt{4\pi}m_{Z'}}\tilde{g}\tilde{Y}_{fL},\quad
\bar{r}_f=\frac{m_Z}{\sqrt{4\pi}m_{Z'}}\tilde{g}\tilde{Y}(f_R),\quad
\bar{\phi}=\frac{m_Z}{\sqrt{4\pi}m_{Z'}}\tilde{g}\tilde{Y}_\phi,
\]
as well as the couplings to the vector and axial-vector fermion
currents, $\bar{v}_f$ and $\bar{a}_f$
\[
\bar{a}_f=\frac{m_Z}{\sqrt{4\pi}m_{Z'}}\tilde{g}a_f,\quad
\bar{v}_f=\frac{m_Z}{\sqrt{4\pi}m_{Z'}}\tilde{g}v_f.
\]
Then, the $Z'$-induced contact couplings are just the quadratic
combinations of  $\bar{l}_f$, $\bar{r}_f$ (or, alternatively,
$\bar{a}_f$ and  $\bar{v}_f$), and the $Z$-$Z'$ mixing effects are
described by $\bar{l}_f\bar{\phi}$, $\bar{r}_f\bar{\phi}$ (or
$\bar{a}_f\bar{\phi}$, $\bar{v}_f\bar{\phi}$).

The above introduced coupling constants are related to the
traditional four-fermion contact couplings (usually marked as
$\epsilon$) through the expressions
\[
\epsilon_{LL} = -\bar{l}_e\bar{l}_f m_Z^{-2}/4, \quad
\epsilon_{RR} = -\bar{r}_e\bar{r}_f m_Z^{-2}/4, \quad
\epsilon_{LR} = -\bar{l}_e\bar{r}_f m_Z^{-2}/4, \ldots
\]
and
\[
\epsilon_{VV} = -\bar{v}_e\bar{v}_f m_Z^{-2}/4, \quad
\epsilon_{AA} = -\bar{a}_e\bar{a}_f m_Z^{-2}/4, \quad
\epsilon_{AV} = -\bar{a}_e\bar{v}_f m_Z^{-2}/4, \ldots
\]
Finally, the $Z$-$Z'$ mixing angle is determined by $\bar\phi$ as
follows,
\[
\theta_0\simeq\frac{m_W
\sin\theta_W}{\sqrt{\alpha}m_{Z'}}\bar\phi,
\]
where $\alpha$ is the fine structure constant.

The leading effects in the cross-section are originated from the
interference of the SM amplitude with the new physics amplitude.
For the Bhabha process, the generic $Z'$-inspired deviation of the
cross-section from its SM prediction is
\begin{equation}\label{dcs-ch}
\Delta d\sigma^\mathrm{Bhabha}/dz = {\cal F}_L \bar{l}_e^2 + {\cal
F}_R \bar{r}_e^2 + {\cal F}_{LR} \bar{l}_e\bar{r}_e +{\cal
F}_{L\phi} \bar{l}_e\bar\phi +{\cal F}_{R\phi} \bar{r}_e\bar\phi.
\end{equation}
where $z$ is the cosine of the scattering angle, and ${\cal
F}={\cal F}(\sqrt{s},z)$ are known functions of the center-of-mass
energy and the scattering angle. It can be also expressed in terms
of the vector and axial vector couplings:
\[
\Delta d\sigma^\mathrm{Bhabha}/dz = {\cal F}_v \bar{v}_e^2 + {\cal
F}_a \bar{a}_e^2 + {\cal F}_{av} \bar{a}_e\bar{v}_e +{\cal
F}_{v\phi} \bar{v}_e\bar\phi +{\cal F}_{a\phi} \bar{a}_e\bar\phi.
\]

For a fixed $z$ the factors ${\cal F}(\sqrt{s},z)$ are smooth
functions of $\sqrt{s}$. However, each factor ${\cal
F}(\sqrt{s},z)$ grows infinitely at $z \to 1$. This is caused by
the photon exchange in the $t$-channel. Due to this singular
behavior the experimental values and the uncertainties change
significantly with increasing of $z$. As it was shown in
\cite{PRD04}, it is possible to make the factors finite for all
the values of the scattering angle by dividing the differential
cross-section by some known monotonic function. For instance, the
factor ${\cal F}_{v}$ is a positive monotonic function of $z$. So
it can be chosen for normalization of the differential
cross-section:
\[
{\cal F}_v^{-1}\Delta d\sigma^\mathrm{Bhabha}/dz = F_L \bar{l}_e^2
+ F_R \bar{r}_e^2 + F_{LR} \bar{l}_e\bar{r}_e +F_{L\phi}
\bar{l}_e\bar\phi +F_{R\phi} \bar{r}_e\bar\phi,
\]
\[
{\cal F}_v^{-1}\Delta d\sigma^\mathrm{Bhabha}/dz = \bar{v}_e^2 +
F_a \bar{a}_e^2 + F_{av} \bar{a}_e\bar{v}_e +F_{v\phi}
\bar{v}_e\bar\phi +F_{a\phi} \bar{a}_e\bar\phi.
\]

This normalization gives us two benefits. First, the obtained
factors $F(\sqrt{s},z)$ are finite for all values of the
scattering angle $z$. Second, the experimental errors for
different bins become equalized that provides the statistical
equivalence of different bins. The latter is important for the
construction of integrated cross-sections.

\section{One parametric fit for the Chiral $Z'$}

The Chiral $Z'$ boson does not interact with the right-handed
species. So, only the factors ${\cal F}_L$ and ${\cal F}_{L\phi}$
survive in Eq. (\ref{dcs-ch}). They are plotted in Fig. 1 for the
center-of-mass energy 200 GeV.
\begin{figure}\label{fig-dcs-ch}
\centering
\includegraphics[bb= 0 0 300 200,width=.5\textwidth]
{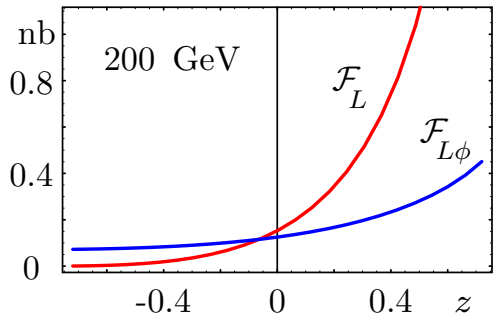} \caption{The factors ${\cal F}_L$ and ${\cal
F}_{L\phi}$ describing the Chiral $Z'$ effects in the differential
cross-section at $\sqrt{s}=200$ GeV.}
\end{figure}
As it was mentioned above, the factors are singular at $z\to 1$. On the
other hand, the normalized deviation of the differential
cross-section from its SM prediction,
\[
{\cal F}_v^{-1}\Delta d\sigma^\mathrm{Bhabha}/dz = F_L \bar{l}_e^2
+ F_{L\phi} \bar{l}_e\bar\phi,
\]
is determined by two finite factors, $F_L$ and $F_{L\phi}$, which
are shown in Fig. 2.
\begin{figure}\label{fig-dcs-ch-norm}
\centering
\includegraphics[bb= 0 0 300 200,width=.5\textwidth]{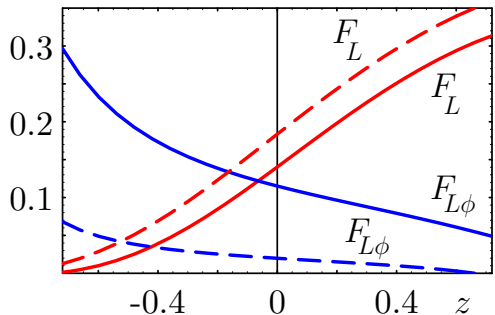}
\caption{ The normalized factors $F_L$ and $F_{L\phi}$ describing
the Chiral $Z'$ effects in the Bhabha process at $\sqrt{s}=200$
GeV (solid lines)  and $\sqrt{s}=500$ GeV (dashed lines).}
\end{figure}

As it is seen, the four-fermion contact coupling $\bar{l}_e^2$
contributes mainly to the forward scattering angles, whereas the
$Z$-$Z'$ mixing term affects the backward angles. At the LEP
energies they can be of the same order of magnitude. The
contribution of the mixing vanishes with the energy growth.

The key problem of the data treatment is the number of independent
parameters which should be determined in an experiment. A large
number of them disperses the experimental statistics leading to
significant uncertainties. This difficulty usually prompts one to
consider various restricted models, supposing some couplings to be
zero. In those cases the loss of generality is the price of the
statistical adequacy of fits. The effects of the Chiral $Z'$ boson
in the Bhabha process are described by two unknown parameters,
only. So, we have a possibility to derive the effective
experimental constraints on them without any additional
restrictions.

First, let us construct a one-parametric observable which is most
preferred by the statistical treatment of data. As is clear, it is
impossible to separate the couplings $\bar{l}_e^2$ and
$\bar{l}_e\bar\phi$ in any observable which is an integrated
cross-section over some interval of $z$. However, the mixing
contribution can be eliminated in the cross-section of the form
(which is inspired by the forward-backward asymmetry)
\[
\Delta\sigma(z^*) = \int_{z^*}^{z_\mathrm{max}} {\cal
F}_v^{-1}(z)\Delta\frac{d\sigma}{dz} \, dz -
\int_{-z_\mathrm{max}}^{z^*} {\cal
F}_v^{-1}(z)\Delta\frac{d\sigma}{dz} \, dz,
\]
where the boundary value $z^*$ should be chosen to suppress the
coefficient at $\bar{l}_e\bar\phi$. The maximal value of the
scattering angle $z_\mathrm{max}$ is determined by a particular
experiment. In this way we introduce the one-parametric
sign-definite observable sensitive to $\bar{l}_e^2$.

The LEP Collaborations DELPHI and L3 measured the differential
cross-sections with $z_\mathrm{max}=0.72$ \cite{diffcs}. The set
of boundary angles $z^*$ as well as the theoretic and experimental
values of the observable are collected in Table 1. The other LEP
Collaborations -- ALEPH and OPAL -- used $z_\mathrm{max}=0.9$
\cite{diffcs}. The corresponding data are presented in Table 2.
\begin{table}\label{tab-1}
\centering \caption{The boundary angles $z^*$ and the theoretic
values of the observable $\Delta\sigma(z^*)$ at
$z_\mathrm{max}=0.72$. The experimental values of the observable
are marked by (D) for DELPHI Collaboration and by (L) for L3
Collaboration.} \rule{0pt}{10pt}\\
\begin{tabular}{|c|c|c|r|} \hline $\sqrt{s}$,
GeV & $z^*$ & $\Delta\sigma(z^*)$ &
$\Delta\sigma^\mathrm{ex}(z^*)$ \\
\hline
 183 & -0.245 & 1742 $\bar{l}_e^2$ & $-2.38 \pm 7.03$ (L)\\
 189 & -0.252 & 1775 $\bar{l}_e^2$ & $-4.28 \pm 3.36$ (D)\\
     & -0.251 & 1771 $\bar{l}_e^2$ & $3.05 \pm 3.65$ (L)\\
 192 & -0.255 & 1788 $\bar{l}_e^2$ & $4.57 \pm 7.32$ (D)\\
 196 & -0.259 & 1806 $\bar{l}_e^2$ & $3.77 \pm 4.33$ (D)\\
 200 & -0.263 & 1823 $\bar{l}_e^2$ & $-1.95 \pm 4.05$ (D)\\
 202 & -0.265 & 1831 $\bar{l}_e^2$ & $1.31 \pm 5.54$ (D)\\
 205 & -0.267 & 1843 $\bar{l}_e^2$ & $-4.09 \pm 3.89$ (D)\\
 207 & -0.269 & 1851 $\bar{l}_e^2$ & $0.40 \pm 3.33$ (D)\\
\hline
\end{tabular}
\end{table}
\begin{table}\label{tab-2}
\centering \caption{The boundary angles $z^*$ and the theoretic
values of the observable $\Delta\sigma(z^*)$ at
$z_\mathrm{max}=0.9$. The experimental values of the observable
are marked by (A) for ALEPH Collaboration and by (O) for OPAL
Collaboration.}\rule{0pt}{10pt}\\
\begin{tabular}{|c|c|c|r|}
\hline $\sqrt{s}$, GeV & $z^*$ & $\Delta\sigma(z^*)$ &
$\Delta\sigma^\mathrm{ex}(z^*)$ \\
\hline
 130 & -0.217 & 2017 $\bar{l}_e^2$ & $-12.40 \pm 19.24$ (A) \\
 &&& $-4.13 \pm 29.29$ (O) \\
 136 & -0.266 & 2092 $\bar{l}_e^2$ & $-50.21 \pm 16.64$ (A)\\
 &&& $-34.18 \pm 31.58$ (O)\\
 161 & -0.370 & 2311 $\bar{l}_e^2$ & $-15.90 \pm 13.24$ (A)\\
 &&& $-14.02 \pm 22.32$ (O)\\
 172 & -0.400 & 2398 $\bar{l}_e^2$ & $-12.11 \pm 12.50$ (A)\\
 &&& $13.71 \pm 17.84$ (O)\\
 183 & -0.424 & 2474 $\bar{l}_e^2$ & $-1.51 \pm ~5.18$ (A)\\
 &&& $11.04 \pm ~5.57$ (O)\\
 189 & -0.435 & 2512 $\bar{l}_e^2$ & $-0.63 \pm ~3.28$ (O) \\
 192 & -0.441 & 2531 $\bar{l}_e^2$ & $-3.48 \pm 9.85$ (O)\\
 196 & -0.447 & 2554 $\bar{l}_e^2$ & $2.96 \pm ~5.09$ (O)\\
 200 & -0.454 & 2577 $\bar{l}_e^2$ & $0.35 \pm ~4.68$ (O)\\
 202 & -0.457 & 2587 $\bar{l}_e^2$ & $-2.87 \pm ~9.00$ (O)\\
 205 & -0.461 & 2604 $\bar{l}_e^2$ & $5.88 \pm ~4.67$ (O)\\
 207 & -0.464 & 2614 $\bar{l}_e^2$ & $-1.42 \pm ~3.46$ (O)\\
\hline
\end{tabular}
\end{table}

The standard $\chi^2$-fit gives the following constraints for the
coupling $\bar{l}_e^2$ at the 68\% CL:
\begin{eqnarray}
  \mathrm{ALEPH:} && \bar{l}_e^2= -0.00304\pm 0.00176 \nonumber\\
  \mathrm{DELPHI:} && \bar{l}_e^2= -0.00054\pm 0.00086\nonumber\\
  \mathrm{L3:} && \bar{l}_e^2= 0.00109\pm 0.00184 \nonumber\\
  \mathrm{OPAL:} && \bar{l}_e^2= 0.00051\pm 0.00064\nonumber\\
  \mathrm{Combined:} && \bar{l}_e^2= -0.00004\pm 0.00048\nonumber
\end{eqnarray}
Hence it is seen that the most precise data of DELPHI and OPAL
collaborations give no signal of the Chiral $Z'$ at the 1$\sigma$
CL. The combined value also shows no signal at the 1$\sigma$ CL.
From the combined fit the 95\% CL bound on the value of
$\bar{l}_e^2$ can be derived, $\bar{l}_e^2<9\times 10^{-4}$.
Supposing the $Z'$ coupling constant $\tilde{g}$ to be of the
order of the electroweak one, $\tilde{g}\simeq 0.6$, the
corresponding $Z'$ mass has to be larger than $0.5$ TeV.

\section{Two parametric fit for the Chiral $Z'$}

Now, let us carry out a complete two parametric fit of
experimental data based directly on the differential
cross-sections. Since in case of the Chiral $Z'$ there are only
two independent couplings, one has to expect that this fit has to
be reliable.

In the fitting we used the available final data for the
differential cross-sections of the Bhabha process. The data set
consists of 299 bins including the data of ALEPH at 130-183 GeV,
DELPHI at 189-207 GeV, L3 at 183-189 GeV, and OPAL at 130-207 GeV
\cite{diffcs}.

The $\chi^2$ function reads
\[
\chi^2(\bar{l}_e,\bar\phi)=\sum_i \left(\frac{\sigma^\mathrm{ex}_i
-
\sigma^\mathrm{th}_i(\bar{l}_e,\bar\phi)}{\delta\sigma^\mathrm{ex}_i}\right)^2,
\]
where $\sigma^\mathrm{ex}_i$ and $\delta\sigma^\mathrm{ex}_i$ are
the measured deviation from the SM value of the differential
cross-section for the $i$th bin (we used the SM predictions, given
by the collaborations) with the corresponding error, and
$\sigma^\mathrm{th}_i$ is the theoretical prediction for the
deviation from the SM due to the Chiral $Z'$ effects. The sum runs
over all the bins.

According to Eq. (\ref{dcs-ch}), the theoretic predictions
$\sigma^\mathrm{th}_i$ are linear combinations of two products of
$Z'$ couplings
\begin{equation}\label{10}
\sigma^\mathrm{th}_i=\sum_{j=1}^{2} C_{ij}A_j,\quad
 A_j=\{\bar{l}^2,\bar{l}_e\bar{\phi}\},
\end{equation}
where $C_{ij}$ are known coefficients. Introducing matrix
notations $\sigma^\mathrm{th}=\sigma^\mathrm{th}_i$,
$\sigma^\mathrm{ex}=\sigma^\mathrm{ex}_i$, $C=C_{ij}$, $A=A_j$,
the $\chi^2$-function can be rewritten as follows
\begin{equation}\label{11}
\chi^2(A)=(\sigma^\mathrm{ex}-CA)^\mathrm{T} D^{-1}
(\sigma^\mathrm{ex}-CA),
\end{equation}
where upperscript T denotes the matrix transposition, and $D$ is
the covariance matrix. The diagonal elements of $D$ are the
experimental errors squared,
$D_{ii}=(\delta\sigma^\mathrm{ex}_i)^2$, whereas the non-diagonal
elements of $D$ are responsible for possible correlations of the
observables.

The $\chi^2$-function has a minimum, $\chi^2_\mathrm{min}$, at the
maximum-likelihood values of the $Z'$ couplings,
\begin{equation}\label{12}
\hat{A}=(C^\mathrm{T}D^{-1}C)^{-1}C^\mathrm{T}D^{-1}\sigma^\mathrm{ex}.
\end{equation}
From Eqs. (\ref{11}), (\ref{12}) we obtain
\begin{eqnarray}\label{13}
\chi^2(A)-\chi^2_\mathrm{min}&=& (\hat{A}-A)^\mathrm{T}
\hat{D}^{-1}(\hat{A}-A),
\nonumber\\
\hat{D}&=& (C^\mathrm{T}D^{-1}C)^{-1}.
\end{eqnarray}

Usually, the experimental values $\sigma^\mathrm{ex}$ are
normal-distributed quantities with the mean values
$\sigma^\mathrm{th}$ and the covariance matrix $D$. The quantities
$\hat{A}$, being the superposition of $\sigma^\mathrm{ex}$, also
have the same distribution. It is easy to show that $\hat{A}$ has
the mean values $A$ and the covariance matrix $\hat{D}$.

The inverse covariance matrix $\hat{D}^{-1}$ is the symmetric
$2\times 2$ matrix, which can be diagonalized. The number of
non-zero eigenvalues is determined by the rank of $\hat{D}^{-1}$,
which equals to the number of linear-independent terms in the
differential cross-section $\sigma^\mathrm{th}$. In the case of
Chiral $Z'$-boson, the rank of $\hat{D}^{-1}$ equals to 2. So, the
right-hand-side of Eq. (\ref{13}) is a quantity distributed as
$\chi^2$ with 2 degrees of freedom (d.o.f.). Since this random
value is independent of $A$, the confidence area in the parameter
space $A=\{\bar{l}_e$, $\bar{\phi}\}$ corresponding to the
probability $\beta$ can be defined as \cite{stat}:
\begin{equation}\label{14}
\chi^2\le \chi^2_\mathrm{min}+\chi^2_{\mathrm{CL},\beta},
\end{equation}
where $\chi^2_{\mathrm{CL,\beta}}$ is the $\beta$-level of the
$\chi^2$-distribution with 2 d.o.f.

The parameter space of the Chiral $Z'$ is the plane ($\bar{l}_e$,
$\bar\phi$). The minimum of the $\chi^2$-function,
$\chi^2_\mathrm{min}=237.29$, is reached at zero value of
$\bar{l}_e$ ($\simeq 10^{-4}$) and almost independent of the value
of $\bar\phi$ (the maximal-likelihood values of the couplings).
The 95\% CL area ($\chi^2_\mathrm{CL}=5.99$) is shown in Figure 3.
\begin{figure}\label{fig-ch-many}
\centering
\includegraphics[bb= 0 0 275 235,width=.5\textwidth]
{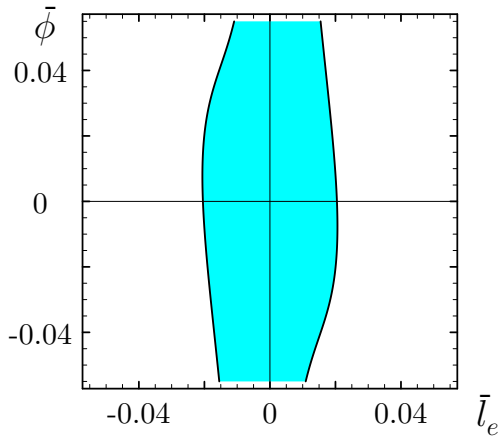} \caption{The 95\% CL area in $\bar{l}_e-\bar\phi$
plane. The final data of ALEPH 130-183 GeV, DELPHI GeV 189-207, L3
183-189 GeV, and OPAL 130-207 GeV are combined.}
\end{figure}

As one can see, the zero point, $\bar{l}_e = \bar\phi= 0$ (the
absence of the Chiral $Z'$ boson) is inside the confidence area.
The value of $\chi^2$ in this point (238.62) is indistinguishable
from the $\chi^2_\mathrm{min}$. In other words, the set of
experimental data cannot determine the signal of the Chiral
$Z'$-boson.

As is seen from Figure 3, the value of $\bar{l}_e$ is constrained
as $\bar{l}_e<0.02$ at the 95\% CL. This upper bound is in an
agreement with the corresponding result of one-parameter fit
($\bar{l}_e<0.03$). Thus, the $Z'$ mass has to be larger than
$0.75$ TeV, if the $Z'$ coupling constant $\tilde{g}$ is again
supposed to be of the order of the electroweak one,
$\tilde{g}\simeq 0.6$.

The fit of the differential cross-sections leads to the better
accuracy on $\bar{l}_e$ than the fit of the integrated
cross-sections based on {\it the same data}. Due to a few number
of independent parameters the increase of the dispersion
(inevitable in case if an extra parameter is added) is compensated
by the increase in the bulk of available data. Without accounting
for the model-independent relations between the $Z'$ couplings it
is impossible to obtain such results.

\section{Discussion}

Let us summarize the results of our investigation. The LEP2 data
for the Bhabha process are adjusted to no signals of the Chiral
$Z'$ both for the one-parameter fit and for the two-parameter one.

We stress once again that the key point of our consideration is
the RG relations between the low energy couplings of the $Z'$ with
the SM fermions. They were systematically used through the
analysis. Therefore only $Z'$ virtual states are responsible for
the deviations of the investigated cross-sections from the SM
values. This is the main feature of the developed approach. If the
relations were not accounted for, no significant constraints have
been found. In contrast, in the ``helicity model fits'' applied by
the LEP collaborations \cite{EWWG} the quantum numbers of a
specific virtual state, which could be responsible for the signal,
remains unknown basically. The such type fits are based on the
effective Lagrangian of the fermion contact interactions. They are
mainly intended to detect the signals of any type virtual states
contributing to a specific ``helicity model'' with one non-zero
contact coupling (AA-model, VV-model, etc.). The more general
four-parametric analysis based on the helicity models of the
fermion contact interactions was carried out in Ref. \cite{pankov}
with the aim to describe the all possible deviations from the SM.
The permissible domains in the parametric space of the models have
been derived. But, in this analysis it is also in principle
impossible to distinguish the specific states responsible for the
deviations. This is because, as it follows from the above
analysis, either the RG relations or the kinematics features of
specific processes were not accounted for.

At LEP1 experiments \cite{LEP1} carried out at $Z$-boson peak the
$Z$-boson coupling constants $g_V$, $g_A$ are precisely measured.
The Bhabha process shows the 1$\sigma$ deviation from the SM
values for Higgs boson masses $m_H\ge 114$ GeV. It is interesting
to estimate the bounds on the $Z$--$Z'$ mixing following from
these experiments. To do that, let us express the measured
parameters $g_V$, $g_A$ through $\bar{l}_e$, $\bar{\phi}$,
\begin{equation}\label{lep1}
g_V-g_V^{\mathrm{SM}}= g_A-g_A^{\mathrm{SM}} = 12.2647
\bar{l}_e\bar{\phi},
\end{equation}
and assume that a total deviation of theory from experiments
follows due to the $Z$--$Z'$ mixing. This gives the low bound on
the mixing. If one assumes other possibilities for deviation the
mixing is increased. In this way one can check is this mixing
excluded by the experiments or not.

\begin{figure}\label{fig-lep1}
\centering
\includegraphics[bb= 0 0 275 235,width=.5\textwidth]
{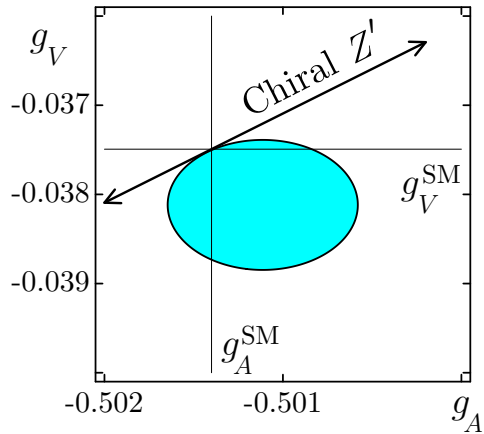} \caption{The 68\% CL area in $g_A-g_V$ plane by the
LEP1 experiments \cite{LEP1}. The SM values of the couplings and
possible deviations of the couplings due to the Chiral $Z'$ are
shown.}
\end{figure}

In Fig. 4 we reproduce the 1$\sigma$ CL area for the Bhabha
process from Ref. \cite{LEP1}. Let us take the SM values of the
couplings corresponding to the top quark mass $m_t=178$ GeV and
the Higgs scalar mass $m_H=114$ GeV. According to Eq.
(\ref{lep1}), the deviation of $g_V$, $g_A$ due to the Chiral $Z'$
boson has to be on the straight line shown in Fig. 4. This line is
completely outside the 1$\sigma$ CL area. So, the Chiral $Z'$ is
excluded in the Bhabha process by the LEP1 experiments. On the
other hand, Fig. 4 demonstrates the 1$\sigma$ deviation from the
SM. The signal could be compatible with another $Z'$ state -- the
Abelian $Z'$ boson. This possibility will be considered in detail
in a separate publication.
%

As it follows from the results of the present paper, there is no
light Chiral $Z'$ gauge boson with the mass $\sim 1$ TeV. Taking
into account the previous investigations \cite{YAF04,PRD04}, we
can conclude that the Abelian $Z'$ boson is the most perspective
neutral vector particle to be searched at the LHC.

This work is supported by the grant F7/296-2001 of the Fundamental
Researches State Fund of Ukraine.

\end{document}